\documentclass[lettersize,journal]{IEEEtran}
\IEEEoverridecommandlockouts
\usepackage{float}
\usepackage{threeparttable}
\usepackage{multirow}
\usepackage{makecell}
\usepackage{xcolor}
\usepackage{amssymb}
\usepackage{cite}
\usepackage{wrapfig}
\usepackage{amsmath,amsfonts}
\usepackage{array}
\usepackage{booktabs} 
\usepackage[caption=false,font=normalsize,labelfont=sf,textfont=sf]{subfig}
\usepackage{textcomp}
\usepackage{stfloats}
\usepackage{url}
\usepackage{verbatim}
\usepackage{graphicx}
\usepackage{algpseudocode}
\usepackage{caption}
\usepackage{amsmath}
\usepackage[ruled,vlined]{algorithm2e}
\usepackage{algpseudocode} 
\usepackage{caption}
\usepackage[dvipsnames]{xcolor}
\usepackage{float}
\usepackage{tabularx}
\usepackage{colortbl}
\definecolor{mygray}{gray}{.9}
\definecolor{mygray1}{gray}{.8}
\definecolor{mygray2}{gray}{.7}
\definecolor{mygray3}{gray}{.6}
\usepackage{diagbox}
\usepackage{enumitem}
\usepackage{tikz}
\usepackage[switch]{lineno,}
\usepackage[dvipsnames]{xcolor}

\newcommand*{\circledd}[1]{\lower.7ex\hbox{\tikz\draw (0pt, 0pt)
   circle (.4em) node {\makebox[0.25em][c]{\small#1}};}}

\usepackage{xcolor}
\definecolor{mygray}{gray}{.9}

\begin{document}

\title{
FlexVector: A SpMM Vector Processor with Flexible VRF for GCNs on Varying-Sparsity Graphs}
\author{~\IEEEmembership{}
{Bohan Li$^{\dag}$}\thanks{$^\dag$These authors contributed equally.}\hspace{0.2cm}
{Shengmin Li$^{\dag}$}\hspace{0.2cm}
{Xinyu Shi}\hspace{0.2cm}
{Enyi Yao}\hspace{0.2cm}
{Francky Catthoor}\hspace{0.2cm}
{Simei Yang$^{*}$\thanks{$^*$Corresponding author.}}
}

\markboth{}
{Shell \MakeLowercase{\textit{et al.}}: A Sample Article Using IEEEtran.cls for IEEE Journals}     

\maketitle

\begin{abstract}

Graph Convolutional Networks (GCNs) are widely adopted for tasks involving relational or graph-structured data and can be formulated as two-stage sparse–dense matrix multiplication (SpMM) during inference. However, existing accelerators often struggle with the irregular workloads induced by power-law node degree distributions. In this work, we propose \textit{FlexVector}, a vector-processor-based architecture that efficiently accelerates SpMM for GCN inference. To address irregular computation patterns, FlexVector adopts a row-wise, product-based dataflow that regularizes SpMM execution and exposes vector parallelism through full-row access to vector registers, eliminating the need for multi-banked register file designs. Building on this dataflow, it introduces software-managed, flexible vector register files (VRFs) that adapt to irregular data access patterns, without sacrificing memory access efficiency. To further exploit these architectural capabilities, we develop a graph-aware preprocessing and node partitioning strategy that restructures irregular graph workloads to better match the row-wise dataflow and VRF capacity. This hardware–software co-design reduces memory traffic, leading to significant performance and energy efficiency gains on real-world GCN workloads. Experimental results on five real-world GCN datasets show that the VRF-centric FlexVector achieves a $3.78{\times}$ speedup and 40.5\% lower energy at comparable area cost relative to a state-of-the-art cache-centric baseline with buffers of the same size.

\end{abstract}


\section{Introduction}
\label{intro}

Graph Convolutional Networks (GCNs) ~\cite{GCN1} have become a widely adopted approach for learning from irregular graph-structured data. By capturing node relationships and connectivity patterns, GCNs can achieve high performance in applications such as social network analysis, traffic prediction, and other graph-based tasks ~\cite{GCN2,GCN3}. The execution of a typical GCN layer is dominated by two primary phases: \textit{aggregation} and \textit{combination}. The aggregation phase depends on the graph’s topology, resulting in sparse and irregular memory access patterns, whereas the combination phase consists primarily of dense operations with regular memory access. 

Due to the contrasting computational characteristics of aggregation and combination, existing GCN accelerators have evolved along two primary architectural directions. The first adopts heterogeneous engines~\cite{hygcn,graphact,sgcn}, which dedicate separate processing engines to the aggregation and combination phases. This enables phase-specific optimizations and streaming execution, where aggregation results directly feed the combination engine for pipelining. However, workload mismatch between the phases can cause pipeline stalls, resource underutilization, and inter-engine synchronization overhead due to rigid partitioning.

The second employs unified engines~\cite{gcnax,awbgcn,grow}, where a single processing engine executes both phases by sequentially mapping them into sparse–dense matrix multiplication (SpMM) operations (Section~\ref{sec:relatedwork}). Such unified designs improve resource utilization and eliminate inter-engine communication. However, executing both phases on a single processing engine makes the architecture highly sensitive to workload irregularity and sparsity variation.

Unified-engine GCN accelerators face challenges due to sparsity arising from three major sources. First, the aggregation phase is much sparser than the combination phase, often by several orders of magnitude~\cite{grow}. Second, nodes exhibit widely varying numbers of neighbors, resulting in highly irregular memory access patterns and computation. Both of these factors largely result from the \textit{power-law distribution} of real-world graphs~\cite{powerlaw1,powerlaw2,powerlaw3} (Section~\ref{sec:relatedwork}), where a few “super-nodes” are densely connected while most nodes have very few connections, particularly affecting the aggregation phase. Third, the sparsity of node features can vary across graphs or over time in evolving datasets. These variations limit the effectiveness of static dataflow optimizations for computation and memory access. To address these challenges, GCNAX~\cite{gcnax} focuses on dataflow optimization for a unified GCN engine, supporting configurable tiling, loop ordering, and loop fusion strategies. In contrast, GROW~\cite{grow} adopts a hardware-software co-design approach, combining a row-wise product dataflow (Gustavson’s algorithm~\cite{gustavson1978two}) with a dedicated hardware cache (or a software-managed scratchpad) and graph preprocessing techniques to handle sparsity and irregular computation.

In this work, we propose a novel unified GCN engine, \textit{FlexVector}, which exploits a vector-processor architecture for SpMM operations and leverages high parallelism and programmability to adapt to varying sparsity. Unlike conventional vector processors, which rely on multi-banked VRFs to sustain access bandwidth, FlexVector employs software-managed \textit{flexible VRFs} with both fixed and dynamic regions to support irregular memory access patterns at the register level. By integrating flexible VRFs with row-wise dataflow in a hardware–software co-designed manner, FlexVector enables full-row VRF access, eliminating the need for VRF banking and simplifying control. Prior work such as GROW~\cite{grow} adopts a cache-centric design with a row-wise dataflow to handle irregular DRAM–cache data access. FlexVector directly applies row-wise dataflow to the buffer–VRF level. However, a single-level dataflow is insufficient for the DRAM–buffer interface in FlexVector’s memory hierarchy, where an additional dataflow (e.g., inner-product) is needed to better align data movement with computation. This motivates FlexVector to adopt a hierarchical dataflow that coordinates row-wise execution at the buffer–VRF level with inner-product execution at the DRAM–buffer level.

Building on this hardware–software co-design, FlexVector incorporates graph-aware preprocessing, similar to prior GCN accelerators~\cite{hygcn,graphact,sgcn,gcnax,grow,gao2023algorithm}. Specifically, the preprocessing partitions input matrices into tiles to address two key challenges. First, the capacity mismatch between large-scale GCN graphs and limited on-chip buffers constrains the amount of data processed at a time. Second, workload imbalance arising from the power-law degree distribution leads to underutilization when buffers are provisioned for high-degree rows. By mitigating these challenges, the preprocessing stage enables efficient utilization of on-chip resources and exploits the benefits of the VRF-centric architecture. We summarize the main \textbf{\textit{contributions}} as follows.

\begin{itemize}
\item \textbf{Hardware Design:} We propose \textit{FlexVector}, a vector-processor-based unified GCN engine featuring software-managed VRFs with fixed and dynamic regions to handle irregular access patterns. We develop a sparsity-aware algorithm to adapt the fixed–dynamic VRF boundary, and introduce a customized coarse-grained ISA to simplify SpMM control.

\item \textbf{Graph Preprocessing:} We introduce a hybrid graph preprocessing strategy that integrates \textit{edge-cut} and \textit{vertex-cut} partitioning to reshape irregular graph workloads under on-chip buffer and VRF capacity constraints. This strategy balances workloads across the sparsity variations of power-law graphs, improving VRF utilization and computation efficiency.

\item \textbf{Dataflow Optimization:} We introduce a hierarchical dataflow, where row-wise product execution at the buffer–VRF level enables full-row VRF access and high lane utilization, while inner-product execution at the DRAM–buffer level sustains efficient data movement across memory hierarchies.
\end{itemize}

We implement FlexVector in RTL and an in-house simulator. We perform a comprehensive PPA evaluation, and conduct ablation studies and sensitivity analyses of buffer and VRF sizing using five real-world graph datasets. Experimental results show that FlexVector achieves up to $3.78\times$ speedup and 40.5\% energy reduction over a state-of-the-art cache-centric baseline at comparable area and buffer capacity. Moreover, the sparsity-aware algorithm that adapts the fixed–dynamic VRF partition achieves near-optimal performance, within 2\% of the best static configuration.

The paper is organized as follows. Section~\ref{sec:relatedwork} reviews SpMM computation in GCNs and relevant related work, highlighting the innovations of our proposed FlexVector. Section~\ref{sec:architecture} introduces the proposed FlexVector architecture. Sections~\ref{sec:graph-preprocess} and \ref{sec:Dataflow} illustrate the graph preprocessing strategy and hierarchical dataflow, respectively. Section~\ref{sec:Experiments} presents experimental evaluations and Section~\ref{sec:conclusion} concludes the paper.

\section{Background and Related Work}
\label{sec:relatedwork}

\subsection{GCN and SpMM Computation}
\subsubsection{GCN Basics} 
\label{subsec:GCNbasics}
\begin{figure}[t]
    \centering
    \includegraphics[width=0.99\linewidth]{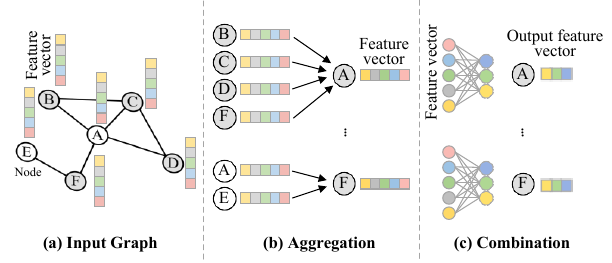}
    \caption{The aggregation and combination stages of GCN.}
    \label{fig:gcn-layer}
    \vspace{-0.5cm}
\end{figure}

Fig.~\ref{fig:gcn-layer} depicts the two main phases of a GCN layer, where \textit{aggregation} propagates information along graph edges and \textit{combination} transforms node features with a dense weight matrix. Equation~\ref{eq:gcn_formula} presents the corresponding forward propagation formulation.

 \begin{equation}
    X^{(l+1)} = \sigma(\hat{A} X^{(l)} W^{(l)})
    \label{eq:gcn_formula}
\end{equation}

\noindent where $X^{(l)}$ is the input feature matrix
 of layer $l$, $W^{(l)}$ is the trainable weight matrix, 
 and $\hat{A}$ denotes the normalized sparse adjacency 
 matrix of the graph structure. $\sigma(\cdot)$ 
 represents a non-linear activation function (e.g., ReLU). 

As reported in prior studies~\cite{awbgcn,gcnax,grow}, matrix $\hat{A}$ is typically large and highly sparse, matrix $X$ has moderate size with workload-dependent sparsity, and matrix $W$ is small and dense. We can compute a GCN layer using two possible execution orders: ($\hat{A}\times X)\times W$ and $\hat{A}\times (X\times W$). The former order involves a sparse–sparse matrix multiplication followed by a dense–dense multiplication, whereas the latter performs two successive sparse–dense multiplications, resulting in significantly fewer multiply-accumulate (MAC) operations compared to the former~\cite{awbgcn,gcnax,grow}. Therefore, we adopt the latter execution order $\hat{A}\times (X\times W$) to perform sparse-dense matrix multiplication (SpMM) for GCN inference.

\subsubsection{Power-Law of GCN Graph}

\begin{wrapfigure}{r}{0.5\columnwidth} 
    \centering
    \includegraphics[width=\linewidth]{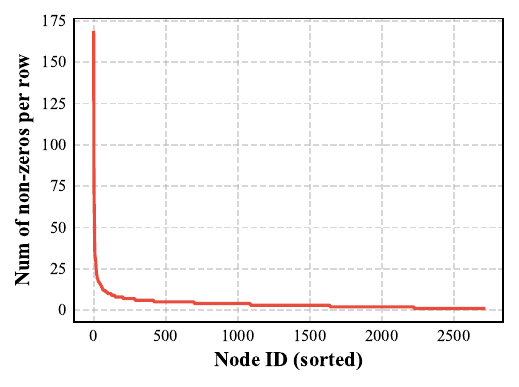}
    \caption{ Power-law distribution of Cora dataset}
    \label{fig:powerlaw-distribution}
    \vspace{-10pt}
\end{wrapfigure}

Fig.~\ref{fig:powerlaw-distribution} illustrates the \textit{power-law distribution} of real-world graphs (e.g., social networks), where a small fraction of ``super-nodes'' have extremely high connectivity, and the vast majority form a ``long tail'' with few neighbors~\cite{powerlaw1,powerlaw2,powerlaw3}. In GCNs, this concentrates non-zero elements in a few adjacency matrix columns, creating structural irregularity that limits memory efficiency in SpMM accelerators. GCN accelerators typically allocate fixed-size on-chip memory (e.g., scratchpad) to optimize data reuse~\cite{grow,awbgcn}. Consequently, super-nodes often overflow fixed memory, causing costly \textit{spill-and-fill} DRAM traffic, while sparse nodes underutilize it. This motivates the use of flexible, cache-like memory combined with graph-aware preprocessing (e.g., GROW~\cite{grow}) to better balance workload. In this work, we also adopt cache-like memory and preprocessing, but our approach differs from GROW~\cite{grow}, as detailed in Table~\ref{tab:comparison} (Section~\ref{sotaGCNAccelerator}).

\subsubsection{SpMM Dataflows}
\label{SPMMdataflows}
\begin{figure}[t]
    \centering
    \includegraphics[width=0.99\linewidth]{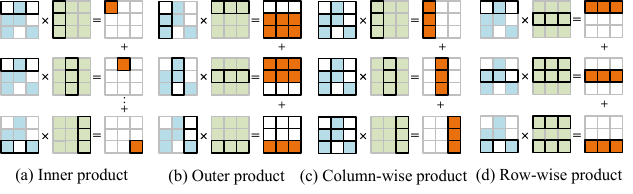}
    \caption{Four typical dataflows for computing SpMM.}
    \label{fig:matrix-multiplication}
    \vspace{-0.5cm}
\end{figure}

Fig.~\ref{fig:matrix-multiplication} illustrates four typical dataflows for computing SpMM, as summarized in~\cite{matraptor,ahsaei2025rethinking}. Each dataflow differs in how it exploits data reuse and its on-chip memory requirements, as follows.

\begin{itemize}
\item \textbf{Inner-product} performs a \textit{row-times-column} computation, producing one output element at a time (Fig.~\ref{fig:matrix-multiplication}(a)). It achieves high reuse of output elements, minimizing the required on-chip accumulation buffer. However, SpMM performance is dominated by irregular memory accesses and index matching.

\item \textbf{Outer-product} performs a \textit{column-times-row} computation, generating partial sums for all output elements (Fig.~\ref{fig:matrix-multiplication}(b)), with high input matrix reuse at the cost of a large accumulation buffer proportional to the output matrix size.

\item \textbf{Column-wise product} performs a \textit{column-times-column} computation, broadcasting a dense column to multiple sparse columns to produce an output column (Fig.~\ref{fig:matrix-multiplication}(c)). It has high reuse of each dense column and requires an accumulation buffer for one output column. However, irregular sparse columns cause workload imbalance in output-column partial-sum computation, increasing control overhead~\cite{awbgcn}.

\item \textbf{Row-wise product} performs a \textit{row-times-row} computation, broadcasting a sparse row to multiple dense rows to produce an output row (Fig.~\ref{fig:matrix-multiplication}(d)). It achieves high reuse of sparse row and requires an accumulation buffer for one output row. However, the column index of each nonzero in the sparse row determines which dense row to fetch, resulting in irregular and repeated dense-row accesses~\cite{grow} (e.g., same dense rows are accessed multiple times when processing different output rows, highlighted in green.)

\end{itemize}

In this work, we adopt \textit{a hierarchical dataflow} that combines \textit{row-wise product} and \textit{inner-product} to optimize data reuse and on-chip memory efficiency. At the buffer–VRF level, we employ the row-wise product to enable full dense-row access within the VRFs, aligning well with vector architectures. While other dataflows can be vectorized, the outer-product requires a larger accumulation buffer, and the column-wise product suffers from severe workload imbalance and control complexity. Although the row-wise product introduces irregular dense-row accesses, it avoids workload imbalance across vector lanes when computing each output row. Additionally, at the DRAM–buffer level, we adopt the inner-product dataflow to maximize reuse of output elements in the buffer.

\subsection{Existing GCN Accelerators}
\label{sotaGCNAccelerator}

\subsubsection{Heterogeneous engines} 

Early designs (e.g., HyGCN~\cite{hygcn}, GraphACT~\cite{graphact}, and SGCN~\cite{sgcn}) adopt heterogeneous engines for sparse–sparse aggregation and dense–dense combination, following the execution order $(\hat{A}\times X)\times W$. These architectures typically pair a SIMD-based aggregation engine with a systolic-array-based combination engine, connected via on-chip buffer/cache to enable streaming execution across phases. However, because the two engines are statically partitioned and exhibit different throughput characteristics, workload imbalance can cause one engine to stall while waiting for the other, leading to pipeline bubbles, reduced resource utilization, and inter-engine synchronization overhead. Additionally, the engines transfer intermediate results via on-chip buffer/cache, and the limited capacity forces designers to partition the graph into smaller tiles (i.e., sub-matrices) that fit in memory. However, such \textit{graph partitioning}~\cite{hygcn,graphact,sgcn} primarily addresses memory constraints and does not fully account for the mismatched computation speeds of heterogeneous engines under irregular workloads.

\subsubsection{Unified engine} 
\label{sec:relatedwork:unified}
Recent GCN accelerators~\cite{awbgcn,gcnax,grow} adopt unified-engine designs to perform successive sparse–dense multiplications, following the execution order $\hat{A}\times (X\times W)$. These works primarily differ in their choice of dataflow. GCNAX~\cite{gcnax} adopts an outer-product dataflow as a basis and explores loop transformations (e.g., tiling, reordering, and fusion) by analyzing the impact on execution cycles and DRAM accesses. In particular, GCNAX performs an exhaustive search over tiling configurations across multiple datasets, evaluating each to find an efficient sub-matrix partitioning. Based on the selected loop transformations, the hardware configures buffer sizes to match the computation and optimize system performance and energy efficiency. Note that outer-product dataflows inherently require larger accumulation buffers than other dataflows (Section~\ref{SPMMdataflows}). In contrast, AWB‑GCN~\cite{awbgcn} employs a column-wise dataflow with multiple processing elements (PEs) connected to a runtime task distributor and queue (TDQ). Instead of using static graph partitions as in prior GCN accelerators~\cite{hygcn,graphact,sgcn,gcnax}, the TDQ dynamically monitors the distribution of nonzero elements and redistributes tasks across PEs at runtime to handle workload imbalance caused by sparsity irregularity. However, this dynamic partitioning incurs additional hardware overhead, including increased routing complexity and control logic (as discussed in Section~\ref{SPMMdataflows} in the context of column-wise dataflow).

\begin{table}[t]
\centering
\caption{Comparison of GROW~\cite{grow} with FlexVector} 
\label{tab:comparison}
\scriptsize
\renewcommand{\arraystretch}{1.25}
\setlength{\tabcolsep}{3pt}

\begin{tabular}{|m{2cm}|m{2.8cm}|m{3.3cm}|}
\hline
& \textbf{GROW~\cite{grow}} & \textbf{FlexVector (our work)} \\ \hline

Architecture &  Unified SpMM engine & Vector-based unified engine\\ \hline

Irregular Handling &  Cache-centric (DRAM$\rightarrow$Cache) & VRF-centric \newline (Buffer$\rightarrow$Registers)\\ \hline

Memory Hierarchy& High-degree node Cache \newline (Hundreds KB, e.g.,512KB) & Multi-buffering (e.g., 2KB) \newline + Flexible VRFs \\ \hline

Preprocessing & Edge-cut (Cache-oriented) & Edge-cut (VRF-oriented) \newline + Vertex-cut \\ \hline

Dataflow & Row-wise (Cache-oriented) & Inner-product (Buffer-oriented) \newline + Row-wise (VRF-oriented) \\ \hline

Control Granularity & Fine-grained \newline(Nonzero × dense row) & Coarse-grained \newline(Sparse row × dense submatrix)\\ \hline

\end{tabular}
\vspace{-0.5cm}
\end{table}

GROW~\cite{grow} is the work most closely related to our proposed FlexVector architecture, as both employ a row-wise product–based dataflow in a unified engine. While this dataflow naturally balances workloads across PEs, it also results in irregular and repeated dense row accesses (see Section~\ref{SPMMdataflows}). Table~\ref{tab:comparison} summarizes the key differences between GROW~\cite{grow} and FlexVector. On the one hand, GROW is a unified SpMM engine which employs a \textit{cache-centric} architecture to handle sparse irregularity. It uses a cache to store precomputed high-degree nodes (HDNs) and employs a \textit{run-ahead} mechanism that leverages this information to execute available dense rows while waiting for others to load, thereby hiding DRAM-cache latency from irregular dense-row accesses. To optimize data locality, it partitions the graph into clusters using an edge-cut preprocessing strategy, processing them sequentially to extract HDN indices and maximize reuse. Nevertheless, GROW’s efficiency heavily depends on preprocessed information buffered in hardware and the availability of a large cache (typically hundreds of KB, e.g., 512KB) to achieve high performance.

On the other hand, our proposed FlexVector leverages a vector-based unified engine to achieve high computational parallelism and programmability. Unlike GROW~\cite{grow}, which relies on a cache-/buffer-centric mechanism to handle irregular accesses, FlexVector adopts a \textit{VRF-centric} approach, shifting potentially repeated accesses from the DRAM–cache/buffer level to the buffer–register interface using flexible, cache-like VRFs. FlexVector combines these VRFs with multi-buffering to hide DRAM access latency and improve overall memory efficiency. Although multi-buffering adds area and energy overhead, a typical FlexVector multi-buffer uses only 2KB (e.g., six-buffering, Section~\ref{sec:FVDeafultConfig}), much smaller than GROW’s hundreds-of-KB caches (e.g., 512KB). In addition, FlexVector applies a hybrid edge- and vertex-cut preprocessing strategy. Edge-cut partitioning creates VRF-oriented graph tiles, where each tile is designed to fit within the VRF capacity rather than the cache/buffer capacity as in GROW. Vertex-cut partitioning then balances workloads within each tile by distributing nonzeros more evenly across sparse rows, addressing imbalance caused by the power-law degree distribution (Section~\ref{sec:graph-preprocess}). Moreover, FlexVector adopts a hierarchical dataflow, performing row-wise computation at the buffer–VRF level to maximize lane utilization and inner-product computation at the DRAM–buffer level to reduce on-chip memory usage for partial sums (Section~\ref{sec:Dataflow}). Finally, FlexVector employs a coarse-grained ISA that manages data access and computation at the granularity of a sparse row multiplying a dense submatrix (e.g., 16×16 elements), whereas GROW uses fine-grained control at the level of a nonzero sparse element multiplying a dense row. FlexVector’s coarse-grained ISA simplifies instruction scheduling by decoupling data movement from computation, trading some control flexibility for simpler scheduling and a reduced instruction count.

\subsection{Vector Processor for SpMM Acceleration} 

\begin{figure*}[t]
    \centering
    \includegraphics[width=0.99\textwidth]{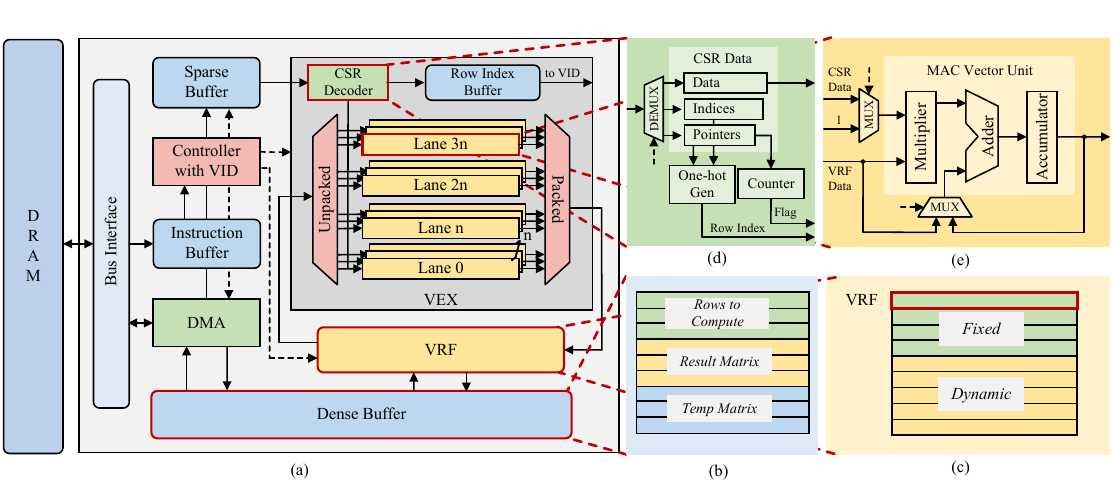}
    \vspace{-0.3cm}
    \caption{Overall architecture of FlexVector}
    \vspace{-0.5cm}
    \label{fig:arch_overview}
    
\end{figure*}

Conventional vector processors implementing the vector-extension~\cite{ara,vicuna,spmvLongVector} offer high SIMD parallelism for matrix kernels, yet still face challenges when performing SpMM in GCN inference. The work of~\cite{spmvLongVector} adopts an inner/outer-product dataflow and relies on gather/scatter ISA and masking to handle irregular accesses, resulting in reduced lane utilization. In contrast, by exploiting a row-wise product dataflow, our FlexVector accesses complete VRF rows across vector lanes, avoiding per-lane indexed accesses. This design eliminates the need for multi-banked VRFs and significantly reduces routing and control complexity.

Similar to FlexVector, IndexMAC~\cite{IndexMAC} also uses a row-wise product dataflow for SpMM computation. It targets structured sparsity, where each matrix block contains a fixed number of nonzero values, and extends the vector ISA with sparse-specific primitives, such as fused index-multiply-accumulate instructions, to improve computation efficiency. These instruction-level optimizations are complementary to FlexVector, with IndexMAC optimized for structured sparsity and FlexVector for fully unstructured sparsity. However, IndexMAC cannot efficiently handle highly irregular patterns, such as power-law distributions.

\section{Proposed FlexVector Architecture}
\label{sec:architecture}

In this section, we detail the microarchitecture and instruction design of FlexVector, a domain-specific vector processor designed as a unified engine for GCN inference.

\subsection{Architecture Overview}
\label{architecture overview}

Fig.~\ref{fig:arch_overview} provides an overview of the FlexVector architecture. The controller, equipped with a Vector Instruction Decoder (VID), reads instructions from the Instruction Buffer and dispatches control signals (dashed lines) to the pipeline components. The Direct Memory Access (DMA) unit loads sparse and dense data from DRAM into the Sparse Buffer and Dense Buffer, respectively. The Vector Execution Unit (VEX) features a CSR (Compressed Sparse Row) decoder for sparse data decoding and configuration, along with multiple parallel computation lanes that access and process data from the vector register files (VRF). Each lane includes a dedicated functional unit to perform row-wise products (see Section~\ref{sec:relatedwork}) for SpMM within the lane. This design allows FlexVector to exploit vector parallelism and programmability for efficient sparse-dense computation in GCN inference.

\subsection{On-Chip Memory Hierarchy}
\subsubsection{On-chip Buffers} 
\label{subsec:OnchipBuffer}
FlexVector employs two types of on-chip buffers. The Sparse Buffer supports sequential streaming access and stores the sparse matrix for SpMM in CSR format. The Dense Buffer stores the dense matrix for SpMM and is logically partitioned into three regions, as depicted in Fig.~\ref{fig:arch_overview}(b): (1) a \textit{Rows-to-Compute region}, buffering input feature rows before they are loaded into the VRF for computation; (2) a \textit{Result Matrix region}, storing the output results of ongoing computations; and (3) a \textit{Temp Matrix region}, customized for FlexVector’s hierarchical dataflow, stores intermediate data for partial sum accumulation during inner-product execution (Section~\ref{sec:innerproductdataflow}). In particular, FlexVector supports multi-buffering in the \textit{Rows-to-Compute} region to hide DRAM–buffer access latency. A typical multi-buffer configuration (e.g., six-buffering) requires only 2\,KB, significantly smaller than the hundreds-of-KB caches (e.g., 512\,KB) used in GROW~\cite{grow} (Section~\ref{sec:FVDeafultConfig}). FlexVector allows adjusting the sizes of Dense Buffer’s regions via compiled instructions (Section~\ref{subsec:coarseISA}) to optimize buffer utilization for different workloads and scheduling strategies. 

\subsubsection{Flexible VRFs}
\label{subsec:FlexibleVRFs}
The flexible VRF mechanism constitutes one of the \textit{\textbf{key innovations}} of the hardware. Like conventional vector processors, FlexVector closely integrates VRFs with its vector computation lanes, enabling high-bandwidth data access. In particular, it introduces software-managed, cache-like \textit{Flexible VRFs} to efficiently handle irregular and sparse data accesses between the VRFs and the lanes. Fig.~\ref{fig:arch_overview}(c) illustrates that the flexible VRFs are logically divided into a fixed region and a dynamic region. The fixed region stores high-reuse rows for SpMM computations in the lanes, while the dynamic region manages low-reuse rows fetched from the Dense Buffer upon VRF misses. The dynamic region operates in a \textit{double-VRF} fashion (Section~\ref{subsec:VRFDataflow}), enabling data fetching to overlap with computation and thereby hiding irregular access latency. This design aligns with the coarse-grained ISA (Section~\ref{subsec:coarseISA}), helping to simplify scheduling.
 Instead of enlarging the VRFs, FlexVector partitions the data movement module to support double-VRF (e.g., VRF depth=16 is split into $8{\times}2$). FlexVector allows the boundary between fixed and dynamic regions to be configured via compiled instructions, enabling adaptation to different sparsity patterns.

\subsection{Vector Execution Units}
\subsubsection{CSR Decoder} 
\label{subsec:CSRDecoder}

Fig.~\ref{fig:arch_overview} (d) illustrates the logic design of the CSR decoder, which takes CSR metadata (pointers, indices, and values) from the Sparse Buffer as input. The internal data path produces two streams: (1) scalar values, which are broadcast to the vector lanes as the multiplicands for SpMM; and (2) row indices, which are converted into a one-hot bitmap and stored in the Row Index Buffer. This bitmap acts as a high-speed address-generation mechanism, enabling the vector lane modules (Section~\ref{subsection:lanes}) to retrieve the corresponding dense rows efficiently. In addition, an internal counter within the CSR decoder tracks the traversal progress of the vector lanes and generates flags that notify them when the accumulation for the current node (i.e., output row) is complete.

\subsubsection{Vector Processing Lanes} 
\label{subsection:lanes}

The VEX comprises multiple parallel lanes that interface with the VRF via configurable unpacked (read) and packed (write) networks (Fig.\ref{fig:arch_overview} (a)), enabling adaptation of the fixed VRF width to the varying precision requirements. For instance, when the VRF width is 128 bits, each lane with a base width of 32 bits can operate at different precisions, such as 8-bit or 32-bit. The unpacking network can divide a VRF entry into four 32-bit elements or sixteen 8-bit elements and distribute them to the corresponding lanes for parallel processing. While this work currently supports 8-bit and 32-bit integer execution, the design could be extended in future work to support other precisions that are powers-of-two. During execution, the lanes receive the broadcast scalar from the CSR decoder and perform row-wise SpMM computation using the multiply-accumulate (MAC) vector unit shown in Fig.~\ref{fig:arch_overview}(e). Upon the CSR decoder signaling completion of the current row accumulation (Section~\ref{subsec:CSRDecoder}), the packed network collects partial results from all lanes, reassembles them into VRF-width words.

\subsection{Coarse-grained ISA}
\label{subsec:coarseISA}

\begin{figure}[t]
    \centering
    \includegraphics[width=0.99\linewidth]{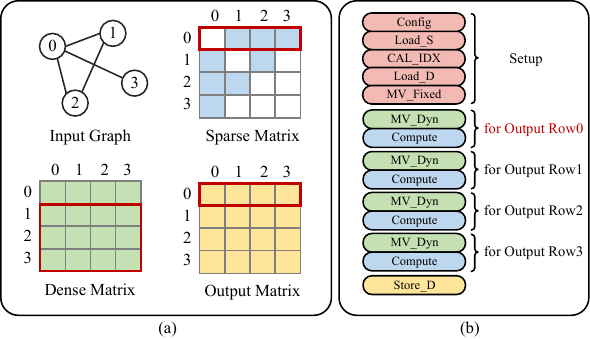}
    \caption{FlexVector's coarse-grained ISA for the SpMM computation of a 4-node graph.}
    \label{fig:coarseISA}
    \vspace{-0.5cm}
\end{figure}

The \textit{coarse-grained} Instruction Set Architecture (ISA) represents another \textit{\textbf{key innovation}} of the FlexVector architecture. It operates at the granularity of a sparse row (or sub-row) paired with a dense matrix (or sub-matrix) stored in the VRFs. In contrast, conventional vector processors~\cite{spmvLongVector} and the prior GCN accelerator GROW~\cite{grow} operate at a finer granularity, processing a single nonzero element together with its corresponding dense row. Fig.~\ref{fig:coarseISA}(a) shows a 4-node graph where the first sparse row represents the connections of node0 and each dense matrix row represents the features of each node. 
Fig.~\ref{fig:coarseISA}(b) presents the coarse-grained ISA used to complete the SpMM computation for this example, and Table~\ref{Table:ISA} summarizes the FlexVector ISA, which consists of two instruction types for row-wise SpMM computation: \textbf{\textit{setup}} and \textbf{\textit{process}}.

\begin{table}[t]
    \caption{Custom ISA of FlexVector}
    \label{Table:ISA}
    \renewcommand\arraystretch{1.4} 
    \scriptsize 
    \centering
    \setlength{\tabcolsep}{1.5pt} 
    
    \begin{threeparttable}
        \begin{tabular}{|m{1cm}<{\centering}|m{1.6cm}<{\centering}|m{5.7cm}<{\centering}|}
            \hline
            \textbf{Type} & \textbf{ISA}  & \textbf{Description} \\ \hline
            
            \multirow{4}{*}{\textbf{Setup}} 
            & Config  & Configure VRF fixed region \\ \cline{2-3}
            & LD\_S / LD\_D  & Load sparse/dense matrix from DRAM \\ \cline{2-3}
            & CAL\_IDX  & Calculate dense row indices of each sparse row\\ \cline{2-3}
            & MV\_Fixed  & Move high-reuse dense rows into the VRF fixed region \\ \hline
            
            \multirow{3}{*}{\textbf{Process}} 
            & MV\_Dyn  & Move dense rows into the VRF dynamic region \\ \cline{2-3}
            & CMP  & Compute output row and write to Dense Buffer \\ \cline{2-3} 
            & ST\_D  & Store output matrix from Dense Buffer to DRAM \\ \hline
            
        \end{tabular}
        \vspace{-0.2cm}
    \end{threeparttable}
\end{table}

\begin{itemize}
    \item \textbf{Setup:} (1) \texttt{Config} sets up the fixed region of the VRFs according to the compiled strategy. (2) \texttt{LD\_S} and \texttt{LD\_D} stream a sparse matrix (or sub-matrix) into the Sparse Buffer and a dense matrix into the Dense Buffer, respectively. (3) \texttt{CAL\_IDX} decodes the CSR data of all sparse rows to compute dense row indices and determine the assignment of dense rows to the VRF fixed region, then generates a one-hot bitmap indicating their positions in the VRFs. This instruction allows run-time computation of indices, eliminating the need for offline precomputation and storage, as required in GROW~\cite{grow}.   
    (4) \texttt{MV\_Fixed} moves high-reuse dense rows from the Dense Buffer into the VRF fixed region.
    
    \item \textbf{Process:} (1) \texttt{MV\_Dyn} uses precomputed row indices (obtained by \texttt{CAL\_IDX}) to move dense rows from the Dense Buffer into the VRF dynamic region. (2) \texttt{CMP} performs SpMM on a sparse row (or sub-row) and a dense matrix (or sub-matrix) in the vector lanes, then writes the resulting row to the Dense Buffer. Since the example graph contains 4 nodes (i.e., 4 sparse rows in Fig.~\ref{fig:coarseISA}(a)), these two instructions are executed 4 times in total (Fig.~\ref{fig:coarseISA}(b)). For the partial-sum case, the \texttt{CMP} instruction sets a flag to accumulate the resulting row with a partial-sum row (loaded by \texttt{MV\_Dyn}) before writing to the Dense Buffer. (3) \texttt{ST\_D} stores the output matrix from the Dense Buffer to DRAM.
    
\end{itemize}

Compared to the fine-grained control in GROW~\cite{grow}, FlexVector adopts a coarse-grained ISA that trades off instruction-level flexibility to simplify scheduling. This is achieved by decoupling data movement (\texttt{MV\_Dyn}) from SpMM computation (\texttt{CMP}) at the granularity of an output row (i.e., a sparse row × dense matrix). By contrast, fine-grained control must interleave data movement and computation for every nonzero in each sparse row. The coarse-grained ISA also reduces instruction count, as quantified in Section~\ref{sec:sensitivity}.

However, this coarse-grained design can lead to workload imbalance across sparse rows due to variations in their number of nonzero elements. Specifically, the VRF must be sized for the worst-case row, i.e., the row with the most nonzero elements, which can lead to underutilization of VRF storage for other rows containing fewer nonzero elements, especially in power-law graph distributions. We mitigate this issue using hybrid preprocessing (Section~\ref{sec:graph-preprocess}) to balance nonzeros across sparse rows and improve VRF utilization.

\section{Hybrid Graph Preprocessing}
\label{sec:graph-preprocess}

This section introduces a \textbf{\textit{hybrid graph-preprocessing strategy}} that restructures SpMM computation to better exploit the FlexVector architecture. The hybrid strategy consists of two steps, \textit{inter-tile edge-cut} and \textit{intra-tile vertex-cut}, designed to address the power-law distribution of GCN graphs and improve memory access efficiency.

\subsection{Inter-tile edge-cut}
\label{subsec:edge-cut}
The \textit{inter-tile edge-cut} strategy partitions the input graph matrix into tiles, with each tile sized to fit the VRFs of the FlexVector architecture. We employ METIS~\cite{metis} to perform inter-tile edge-cut, which minimizes cross-tile edges and keeps highly connected vertices within the same tile to preserve data locality. 
In contrast to typical strategies that partition GCN graphs to fit on-chip buffer sizes ~\cite{hygcn,graphact,sgcn,gcnax,grow}, our approach skips buffer-level partitioning and directly partitions graphs to fit the VRF capacity. This VRF-oriented partitioning strategy prioritizes data locality for row-wise products in the VRFs and aligns with the coarse-grained ISA execution in the vector lanes (Section~\ref{subsec:coarseISA}). We then group multiple tiles in the Dense Buffer to enable buffer-level data reuse (Section~\ref{sec:Dataflow}).

\subsection{Intra-tile vertex-cut}
\label{sec:Intra-tileVertex-cut}
After the inter-tile edge-cut, we aim to further balance the workload across sparse rows within each tile to mitigate the imbalance caused by the power-law degree distribution of GCN graphs. In our evaluation on the \textit{Cora} dataset, tiles (e.g., a 16$\times$16 submatrix) generated by the inter-tile edge-cut exhibit highly imbalanced row densities. We use $RNZ$ to denote the number of nonzero elements per sparse row.
Only about 5\% of rows are densely populated with $RNZ \ge 9$, including extreme cases that occupy the full row width ($RNZ = 16$), whereas the vast majority, approximately 95\%, remain sparse with $RNZ \le 4$. As discussed in Section~\ref{subsec:coarseISA}, FlexVector adopts a coarse-grained ISA that executes SpMM with all required dense rows resident in the VRFs, requiring the VRF depth to be no less than the maximum per-row $RNZ$.

\begin{figure}[t]
    \centering
    \includegraphics[width=0.99\linewidth]{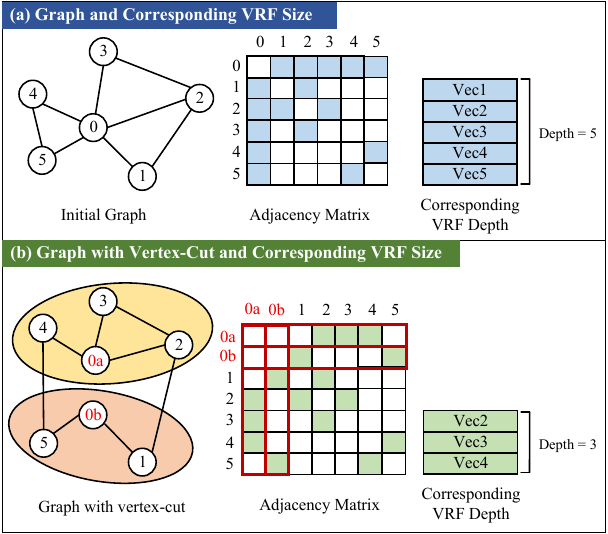}
    \caption{Comparison of VRF provisioning before and after vertex-cut preprocessing. 
    (a) Before vertex-cut, a super-node row has RNZ$=5$. 
    (b) After vertex-cut, the row is split and RNZ$\le3$, reducing VRF depth.}
    \label{fig:vertex-cut}
    \vspace{-0.5cm}
\end{figure}

To address this constraint, we apply an \textit{intra-tile vertex-cut} that splits high-degree vertices and redistributes their associated nonzero elements across multiple sparse rows to balance the workload. Fig.~\ref{fig:vertex-cut} provides an intuitive comparison of the graph structure before and after this optimization. In Fig.~\ref{fig:vertex-cut}(a), Node0 has the highest degree ($RNZ = 5$), implying that the VRF depth for data input must be at least 5. In Fig.~\ref{fig:vertex-cut}(b), after applying the vertex-cut, Node0 is split into nodes $0a$ and $0b$, partitioning its nonzero elements across two sparse rows and reducing the maximum row density to 3, thereby improving workload balance across rows.

Algorithm~\ref{alg:vertex_cut} illustrates our intra-tile vertex-cut strategy, which takes the original tile and a per-row $RNZ$ bound $\tau$ as inputs, generating a new tile in which no row exceeds $\tau$ nonzero elements. Note that $\tau$ defines a preprocessing constraint rather than the actual VRF depth, which can be provisioned higher (Section~\ref{subsec:VRFDataflow}). For rows with $RNZ > \tau$ (e.g., 3 in Fig.~\ref{fig:vertex-cut}(b)), the algorithm generates the $MissList$ and $HitList$ by analyzing the sparse tile. It assumes an ideal VRF depth of $\tau$ and that the dense rows corresponding to the sparse indexes with the most nonzero elements (e.g., sparse indexes 0–2 in Fig.~\ref{fig:vertex-cut}(a)) are already loaded. Under these assumptions, the $MissList$ includes sparse index 3–5, corresponding to dense rows 3–5. The algorithm then computes the required number of splits ($K=2$, $\lceil 5/3 \rceil$ for Node0 in Fig.~\ref{fig:vertex-cut}(a)) and determines the number of misses and hits for each split. Subsequently, the algorithm distributes the misses and hits evenly across the splits, with each split represented by a sub-row. To this end, the algorithm pops $n_\text{miss}$ (e.g.,2) and $n_\text{hit}$ (e.g.,1) elements from the $MissList$ and $HitList$, respectively, and combines them to form a new sub-row (line 11-15). For instance, combining sparse index3-4 from the $MissList$ and index2 from the $HitList$ forms the sub-row of $0a$ in Fig.~\ref{fig:vertex-cut}(b). Finally, after generating all sub-rows, the algorithm outputs a new tile with balanced workloads. Section~\ref{sec:exp:vertex_cut} demonstrates the effectiveness of Algorithm~\ref{alg:vertex_cut} through experimental results.

\begin{algorithm}[t]
\small
\caption{Intra-tile Vertex-Cut Strategy}
\label{alg:vertex_cut}
\LinesNumbered
\SetKwInOut{Input}{Input}
\SetKwInOut{Output}{Output}

\Input{Original sparse tile $T$, per-row $RNZ$ bound $\tau$}
\Output{New sparse tile $T_{new}$ after workload balance}

$T_{new} \gets \emptyset$

\For{$r \in T$}{
    \If{$RNZ \le \tau$}{
        $T_{new} \gets T_{new} \cup r$
    }
    \Else{
        \tcp{Step 1: Separate Miss/Hit indices}
        $MissList, HitList \gets Analyze(T)$

        $K \gets \lceil RNZ / \tau \rceil$ \tcp{Total splits needed}

        $n_{miss} \gets \lceil |MissList| / K \rceil$\;
        $n_{hit} \gets \tau - n_{miss}$

        \tcp{Step 2: Distribute into sub-rows}
        \For{$i \in [0, K-1]$}{
            $r_{sub} \gets CreateSubRow()$\;
            $SubMiss \gets Pop(MissList, n_{miss})$\;
            $SubHit \gets Pop(HitList, n_{hit})$\;
            $r_{sub}.Indices \gets SubMiss \cup SubHit$\;
            $T_{new} \gets T_{new} \cup r_{sub}$
        }
    }
}

\Return{$T_{new}$}
\end{algorithm}

\section{Hierarchical Dataflow}
\label{sec:Dataflow}

This section presents a hierarchical dataflow in which the \textit{row-wise product} operates at the buffer-VRF level, while the \textit{inner-product} operates at the DRAM-buffer level.

\subsection{Buffer-VRF level: Row-wise product dataflow}
\label{subsec:VRFDataflow}

We apply a row-wise dataflow at buffer-VRF level for each tile (sparse matrix $\times$ dense matrix $=$ output matrix) after vertex-cut. FlexVector includes cache-like, flexible VRFs consisting of a \textit{fixed region} and a \textit{dynamic region}, with the boundary between these regions adjustable for each tile according to sparsity variations using compiled configuration instructions (\texttt{Config}). Based on compile-time analysis of sparse tiles, we prioritize placing the top-$k$ most frequently accessed dense rows in the \textit{fixed region} of the VRFs. For tiles with high-reuse rows, $k$ is set as large as possible to maximize reuse, whereas for tiles with low-reuse rows (e.g., used only once), $k$ is reduced, potentially to zero, subject to VRF capacity constraints. The optimized $k$ also depends on whether the VRFs support a double-VRF scheme (i.e., double dynamic regions), which allows overlapping data loading from the Dense Buffer and computation to hide access latency.

Algorithm~\ref{alg:topkfixed} presents the top-$k$ selection strategy for the VRF fixed region. It takes a sparse tile, per-row $RNZ$ bound, a starting fraction to initialize $k$, VRF depth, and VRF mode (e.g., single-VRF, double-VRF) as inputs, and returns the optimized $best\_k$. First, we analyze the sparse tile to compute the number of nonzero elements per column (denoted by $CNZ$) and sort the columns in descending order to form the $Sorted\_CNZ$ list, prioritizing the highest-density columns for selection (line 1). We then set the initial top-$k$ value as a fraction (e.g., $pct=0.5$) of the per-row RNZ bound and iteratively adjust it to ensure the VRF depth can accommodate the selected rows (line 2). In each iteration, we select the current top-$k$ indices from $Sorted\_CNZ$ as candidate fixed-region rows. Based on this top-$k$ selection, we compute the per-row miss counts for the sparse tile and sort them in descending order to form the $Sorted\_Miss$ list, which indicates how many accesses would miss in the VRF if the selected dense rows were fixed (line 5). In single-VRF mode, we consider only the row with the largest miss count. In double-VRF mode, we consider the two rows with the highest miss counts to account for overlapping the computation of the current dense rows (corresponding to the current sparse row) with the loading of dense rows for the next sparse row (lines 6–9). If the VRF depth can accommodate these rows, we increment $k$ and update $best\_k$, and otherwise, we decrement $k$ (line 10-13). The iteration continues until $k$ cannot be further adjusted without exceeding the VRF capacity.

\begin{algorithm}[t]
\small
\caption{Top-$k$ VRF Fixed Region Selection}
\label{alg:topkfixed}
\LinesNumbered
\SetKwInOut{Input}{Input}
\SetKwInOut{Output}{Output}

\Input{Sparse tile $T$, per-row $RNZ$ bound $\tau$, start fraction $pct$, VRF depth $D$, VRF mode $mode$}
\Output{Selected top-$k$ value $best\_k$}

$Sorted\_CNZ \gets SortDesc(Analyze(T))$\;
$k \gets \lceil \tau \times pct \rceil$, $best\_k \gets k$\;

\While{$k > 0 \ \textbf{and} \ k \le D$}{
    $TopkIDX \gets Sorted\_CNZ[0:k-1]$\;
    $Sorted\_Miss \gets SortDesc(Analyze(T, TopkIDX))$\;

    \If{mode = single buffer}{
        $fit \gets k + Sorted\_Miss[0] \le D$\;
    }
    \ElseIf{mode = double buffer}{
        $fit \gets k + Sorted\_Miss[0] + Sorted\_Miss[1] \le D$\;
    }
    
    \If{fit}{ $best\_k \gets k$, $k \gets k + 1$ }
    \Else{ $k \gets k - 1$ }
}

\Return $best\_k$\;
\end{algorithm}

\begin{figure}[t]
    \centering
    \includegraphics[width=0.99\linewidth]{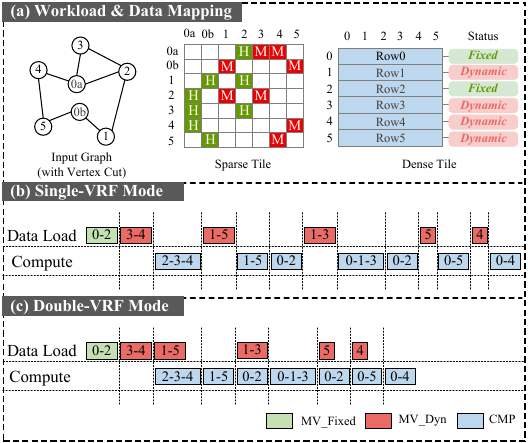}
    \caption{Comparison of Execution Modes. (a) Graph topology maps to Fixed (green) or Dynamic (red) rows based on reuse. (b) Single-VRF Mode: \texttt{MV\_Dyn} execution waits for \texttt{CMP} to complete. (c) Double-VRF Mode: Pipelining allows concurrent \texttt{MV\_Dyn} and \texttt{CMP} to hide data movement overhead.}
    \label{fig:pipeline_comparison}
    \vspace{-0.5cm}
\end{figure}

Fig.~\ref{fig:pipeline_comparison} compares the single-VRF mode with the double-VRF mode. For the example in part (a), the column indexes $0a-0b$ and $2$ of the sparse tile are used to select dense rows in the VRF fixed region. Part (b) shows the execution trace for the single-VRF mode. Since FlexVector uses a coarse-grained ISA (Section~\ref{subsec:coarseISA}) to execute \texttt{MV\_Dyn} and \texttt{CMP} for each sparse row, dense rows cannot be overwritten until the computation completes. In contrast, part (c) shows the double-VRF mode, which includes two dynamic regions and prefetches dense rows for the next sparse row, enabling pipelined execution of data movement and SpMM computation while improving pipeline and lane utilization. As previously discussed in Section~\ref{subsec:FlexibleVRFs}, rather than increasing the VRF capacity to support double-VRF, we partition the data movement module to enable concurrent data transfer and computation. This design achieves higher system performance while having negligible impact on area and energy overhead, with its benefits quantified in Section~\ref{sec:exp:double-vrf}.

 \subsection{DRAM-buffer level: Inner-product dataflow}
\label{sec:innerproductdataflow}

\begin{figure}[t]
    \centering
    \includegraphics[width=0.99\linewidth]{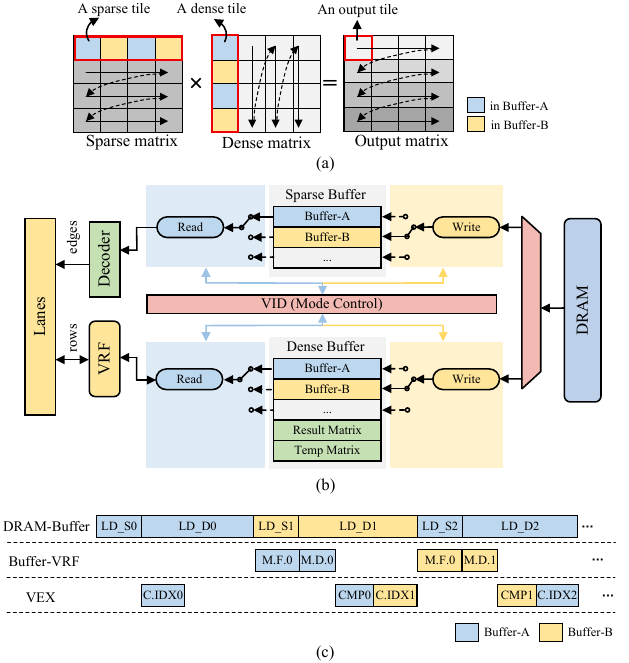}
    \caption{Inner-product dataflow at DRAM-buffer level for each tile. (b) Multi-buffer organization (depicted as dual-buffer A/B). (c) Execution trace showing DRAM transfers overlapping with SpMM computation to hide latency. M.F.: MV\_Fix; M.D.: MV\_Dyn; C.IDX: CAL\_IDX.}
    \label{fig:dataflow-allocation}
    \vspace{-0.5cm}
\end{figure}

Fig.~\ref{fig:dataflow-allocation} illustrates the inner-product dataflow at DRAM-buffer level for each tile, where the size is determined in a \textit{VRF-capacity-aware} manner. The inner product enables output-stationary computation for each output tile in the Dense Buffer before the final result is produced (Fig.~\ref{fig:dataflow-allocation}(a)). The Dense Buffer can hold $m \ge 2$ tiles in the \textit{Result Matrix region} (Section~\ref{subsec:OnchipBuffer}), each with a footprint bounded by the VRF capacity.
The $m$ tiles can be distributed across multiple buffers (e.g., Buffer A, B, …) to overlap data loading from DRAM with computation. For example, $m=2$ corresponds to a conventional double buffer. The Sparse Buffer adopts a similar multi-buffer organization (Fig.~\ref{fig:dataflow-allocation}(b)). FlexVector preloads the dense rows required by each sparse tile into the Dense Buffer before SpMM computation, avoiding repeated DRAM fetches. In contrast, GROW~\cite{grow} performs irregular DRAM accesses whenever a row miss occurs. By shifting repeated irregular accesses to the Dense Buffer–VRF interface, FlexVector reduces DRAM–Buffer misses and improves overall system performance, as demonstrated in Section~\ref{subsec:various_buffer}. Exploration of alternative inter-tile dataflows, such as outer-product, is left for future work and is beyond the scope of this paper.

Fig.~\ref{fig:dataflow-allocation}(c) shows a tile-level execution trace. The trace begins with loading the sparse tile (\texttt{LD\_S}), after which the sparse matrix is decoded via index calculation (\texttt{CAL\_IDX}) executed in parallel with dense matrix loading (\texttt{LD\_D}). The \texttt{MV\_Fixed} and \texttt{MV\_Dyn} instructions then move dense rows from the Dense Buffer to the VRFs, followed by \texttt{CMP} to perform the SpMM computation. The \texttt{MV\_Dyn} and \texttt{CMP} operations execute multiple times in a pipelined manner depending on the number of rows in the tile, as shown in Fig.~\ref{fig:dataflow-allocation}(c) for simplicity. FlexVector then writes the resulting output tile to the Dense Buffer, either to the \textit{Temp Matrix} region if it is an intermediate tile for the inner-product dataflow, where it is subsequently accumulated into the \textit{Result Matrix}, or directly to the \textit{Result Matrix} region if it is the final output tile (Section~\ref{subsec:OnchipBuffer}, Fig.~\ref{fig:arch_overview}). In addition, thanks to the multi-buffering scheme, FlexVector computes one set of tiles while loading the next, overlapping computation with DRAM accesses and hiding memory latency. As discussed in Section~\ref{sec:relatedwork:unified}, although multi-buffering can introduce energy and area overhead, FlexVector uses a small buffer (2KB for six-buffering, Section~\ref{sec:Exp-baseline}), which is significantly smaller than the hundreds-of-KB caches used in GROW~\cite{grow}. This design preserves the benefits of multi-buffering without excessive resource cost, as quantified in Section~\ref{subsec:various_buffer}. 
In summary, the hierarchical dataflow and flexible VRF data-access strategies, together with dedicated hardware designs (Section~\ref{sec:architecture}) and graph preprocessing (Section~\ref{sec:graph-preprocess}), enable FlexVector to maintain efficient memory access and high utilization across the diverse sparsity patterns.

\section{Experiments}
\label{sec:Experiments}
\subsection{Experiment setup}

\subsubsection{Methodology}
We implement the FlexVector processor in SystemVerilog, and synthesize the RTL using Synopsys Design Compiler with a 28nm standard cell library at 1\,GHz to obtain the area and power consumption of each hardware component. We model the area and energy of on-chip SRAM and VRFs using CACTI 7.0~\cite{cacti7}. For off-chip DRAM, we adopt HBM 1.0 with a bandwidth of 128\,GB/s and an energy cost of 7\,pJ/bit~\cite{oconnor2014hbm}. Based on RTL and synthesis results, we develop an in-house instruction-driven simulator in Python to evaluate PPA tradeoffs across different hardware--software design parameters (e.g., SRAM sizes, preprocessing strategies). 

\subsubsection{Datasets}
We conduct evaluations on five representative GCN datasets, covering a wide range of graph scales and sparsity patterns, as summarized in Table~\ref{tab:datasets}. Cora and CiteSeer are small citation graphs with high feature dimensionality, while Pubmed is a medium-scale biomedical citation graph. Reddit and Yelp are large-scale graphs with tens of millions of edges, representing social and business review networks, respectively.

\subsubsection{FlexVector Default Configurations}
\label{sec:FVDeafultConfig}
We define the default design configurations for FlexVector. The Dense Buffer is 2KB, the Sparse Buffer is 256B, and the multiple-buffer mechanism uses $m=6$ to manage the rows-to-compute region (Section~\ref{subsec:OnchipBuffer}). The VRF has 128bit rows, each holding 16 8-bit elements, and a depth of $6{\times}2=12$, corresponding to the double-VRF configuration with a vertex-cut threshold of $\tau=6$ (Section~\ref{sec:Intra-tileVertex-cut}). 

\begin{table}[t]
\centering
\caption{Dataset Specifications.}
\label{tab:datasets}
\renewcommand{\arraystretch}{1.15}
\small
\begin{tabular}{lccc}
\hline
\textbf{Dataset} & \textbf{Nodes} & \textbf{Edges} & \textbf{Feature Dim} \\ \hline
Cora     & 2,708     & 5,429      & 1,433 \\
CiteSeer & 3,327     & 4,732      & 3,703 \\
Pubmed   & 19,717    & 44,338     & 500   \\
Reddit   & 232,965   & 11,606,919 & 602   \\
Yelp     & 716,847   & 13,954,819 & 300   \\ \hline
\end{tabular}
\end{table}

\begin{figure}[t]
    \vspace{-0.4cm}
    \hspace*{0.5cm}
    \includegraphics[width=0.8\linewidth]{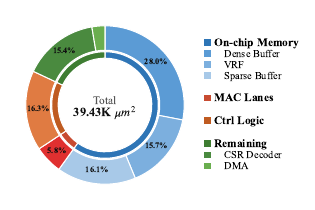}
    \vspace{-0.45cm}
    \caption{Area breakdown of FlexVector (total: 39.43K\,$\mu m^2$).}
    \label{fig:area}
    \vspace{-0.5cm}
\end{figure}

\subsubsection{Baselines}
\label{sec:Exp-baseline}

As discussed in Section~\ref{sec:relatedwork:unified}, GROW~\cite{grow} is the work most closely related to FlexVector. To compare, we construct a GROW-like system that preserves GROW’s key mechanisms~\cite{grow} (Section~\ref{sec:relatedwork:unified}): (1) a cache-centric memory hierarchy that preloads the top-$N$ high-degree node (HDN) dense rows into a given-capacity buffer (e.g., software-controlled cache); (2) a run-ahead execution mechanism that skips stalled rows and continues executing subsequent rows in the buffer (up to a look-ahead depth of 16~\cite{grow}), while missed rows are queued for DRAM fetch; and (3) fine-grained ISA control, where each instruction operates on a single nonzero sparse element and its corresponding dense row. In contrast, FlexVector uses coarse-grained ISA to manage computation and data movement for an entire sparse row multiplied by a dense submatrix (e.g., tile size=$16{\times}16$, Section~\ref{subsec:coarseISA}). We use two GROW-like variants as baselines: (1) \textbf{\textit{GROW-like}}, with small memory configuration similar to the default FlexVector configuration (2KB Dense Buffer and 256B Sparse Buffer, with multi-buffering factor $m=6$), and (2) \textbf{\textit{GROW-like$^\dagger$}}, with the large memory configuration originally used in GROW~\cite{grow} (512KB Dense Buffer and 12KB Sparse Buffer, with $m=2273$).

\subsection{Area Evaluation}
Fig.~\ref{fig:area} shows the area breakdown of FlexVector under the default configuration, with a total area of 39.43K\,$\mu m^2$. On-chip memory dominates at 59.9\% (Dense Buffer 28.0\%, VRF 15.7\%, Sparse Buffer 16.1\%), reflecting the memory-centric nature of GCN inference. MAC Lanes account for 5.8\% of the total area. Control logic, comprising the VEX core control and top-level controller (VID), accounts for 16.3\%. The remaining 18.0\% is occupied by the CSR Decoder and DMA.

\begin{figure*}[t]
    \centering
    \minipage{0.49\linewidth}
        \includegraphics[width=\linewidth]{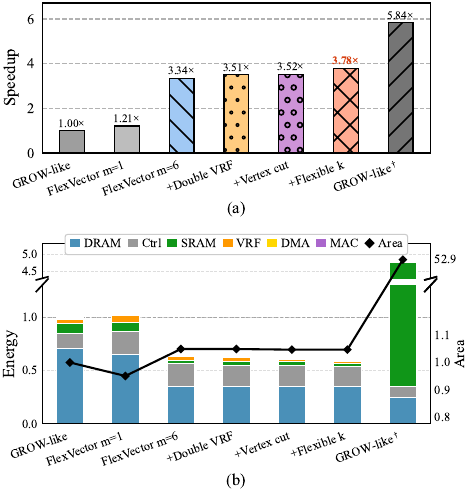}
        \vspace{-0.5cm}
        \label{fig:ablation_multi}
    \endminipage
    \hfill
    \minipage{0.49\linewidth}
        \includegraphics[width=\linewidth]{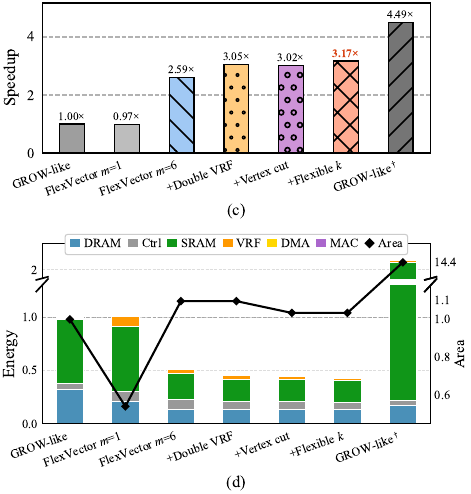}
        \vspace{-0.5cm}
        \label{fig:ablation_single}
    \endminipage
    \caption{Ablation study of FlexVector. (a–b) Speedup, energy, and area averaged across five datasets, normalized to a GROW-like baseline (tile size = 16$\times$16).
    (c–d) Case-study on a representative dataset, Pubmed, evaluating scalability under larger tile configurations (tile size = 64$\times$64), normalized to a GROW-like baseline with the same tile size.}
    \label{fig:ablation}
    \vspace{-0.5cm}
\end{figure*}

\subsection{Ablation Study}
\label{subsec:ablation_study}

Fig.~\ref{fig:ablation} shows the evolution of speedup, energy, and area as optimizations are incrementally applied.
Fig.~\ref{fig:ablation}(a–b) report the geometric mean across five datasets (tile size = 16×16), whereas (c–d) present a case-study on a representative dataset, Pubmed (medium-scale dataset), to evaluate scalability under larger tile configurations (tile size = 64×64). The larger tile configuration in the latter corresponds to increased VRF and buffer sizes (4$\times$ in both memory width and depth relative to the default configuration).

\subsubsection{Impact of Shifting Irregular Dense-row Access to the Buffer–VRF Interface}
FlexVector~($m{=}1$) refers to the design without multiple-buffering support for the Dense and Sparse Buffers. Compared to the GROW-like baseline, which has buffer capacities corresponding to $m{=}6$ (Section~\ref{sec:Exp-baseline}), FlexVector~($m{=}1$) achieves a 1.21$\times$ speedup (Fig.~\ref{fig:ablation}(a)) by confining irregular dense-row accesses from the DRAM--buffer level in GROW-like to the buffer-VRF interface. This performance improvement comes with lower area cost (4.9\% less) and comparable energy consumption (Fig.~\ref{fig:ablation}(b)). For the case-study on larger tile configuration (Fig.~\ref{fig:ablation}(c–d)), FlexVector~($m{=}1$) delivers 3\% lower performance than GROW-like ($m{=}6$), but benefits from a reduced area cost of 46\% due to its smaller buffer sizes. Note that the geometric mean in Fig.~\ref{fig:ablation}(a–b) represents an average across all datasets, whereas performance can vary for individual datasets with different sparsity levels or tile configurations. 

\subsubsection{Impact of Multi-buffering Support}
FlexVector~($m{=}6$) increases the buffer capacity to match that of GROW-like, enabling a multiple-buffer design that hides DRAM access latency. For Fig.~\ref{fig:ablation}(a-b), this configuration achieves a 3.34$\times$ speedup with only a 4.9\% area overhead (due to a more complex controller and the introduction of VRF) compared to GROW-like, while reducing total energy by 36\%. The performance and energy gains come primarily from eliminating repeated irregular DRAM accesses and confining them to the buffer–VRF interface. The case-study in Fig.~\ref{fig:ablation}(c–d) shows similar PPA trends (2.59$\times$ speedup, 9.6\% area overhead, and 48.9\% lower energy vs. GROW-like), confirming that these improvements generalize to larger tile configuration.

\subsubsection{Impact of Double-VRF Support}
\label{sec:exp:double-vrf}
 FlexVector~(+Double VRF) extends FlexVector~($m{=}6$) by partitioning the VRF depth from $16$ into $8{\times}2$ (Fig~\ref{fig:ablation} (a-b)), allowing one half to serve the current computation while the other prefetches the next dense rows. This enables concurrent data movement and SpMM execution under the coarse-grained ISA (Section~\ref{subsec:coarseISA}).  This configuration provides an additional 5.1\% speedup to reach $3.51{\times}$, while reducing energy by 2.1\% owing to the shorter execution time lowering control and leakage energy. Since the total VRF capacity remains unchanged (256\,B), there is no additional area overhead. Under larger tile configurations in Fig.~\ref{fig:ablation}(c–d), FlexVector~(+Double VRF) partitions the VRF depth from $64$ into $32{\times}2$. Compared to FlexVector~($m{=}6$), FlexVector~(+Double VRF) achieves 15\% performance improvement and 10.8\% energy reduction. These gains are more pronounced in this setting because larger tiles provide longer execution windows, enabling more effective overlap between VRF data access and SpMM computation (via MV\_Dyn and CMP coarse-grained ISA, Section~\ref{subsec:coarseISA}).

\subsubsection{Impact of Vertex-cut (Algorithm~\ref{alg:vertex_cut})}
\label{sec:exp:vertex_cut}

FlexVector~(+Vertex cut) balances nonzeros across sparse rows and reduces VRF pressure (Section~\ref{sec:Intra-tileVertex-cut}), lowering the VRF depth from $8{\times}2$ in FlexVector~(+Double VRF) to $6{\times}2$ (Fig~\ref{fig:ablation} (a-b)). This optimization achieves an additional 0.3\% speedup ($3.52{\times}$) while reducing energy by 2.0\% due to smaller VRF capacity. The VRF shrinks from 256\,B to 192\,B, yielding a 3.1\% reduction in VRF area (0.4\% total chip area reduction). Notably, the primary benefit of +Vertex cut lies in reducing VRF area, which scales with tile size. Under larger tile configurations (Fig.~\ref{fig:ablation}(c–d)), the VRF depth is reduced from $32{\times}2$ to $8{\times}2$, achieving an additional 48.2\% reduction in VRF area (5.7\% total area), along with 2.2\%  energy reduction and a 1.5\% latency overhead compared to FlexVector~(+Double VRF). These results suggest that +Vertex cut can be more beneficial under larger tile configurations, where VRF pressure is higher.

\subsubsection{Impact of Introducing a Fixed Region in the VRF}
FlexVector~(+Flexible $k$) introduces a fixed region of $k$ VRF rows (set via Algorithm~\ref{alg:topkfixed}), enabling both fixed and dynamic regions in the VRF (Section~\ref{subsec:VRFDataflow}), whereas FlexVector~(+Vertex cut) uses a fully dynamic VRF. This fixed-dynamic VRF configuration keeps frequently accessed data in the fixed region, reducing buffer–VRF accesses, and delivers an additional 7.4\% speedup ($3.78{\times}$) and 3.1\% energy reduction, with no area increase compared to FlexVector~(+Vertex cut) (Fig.~\ref{fig:ablation}(a–b)). This observation is consistent in Fig.~\ref{fig:ablation}(c–d), where +Flexible $k$ achieves an additional 4.7\% speedup and 3.1\% energy reduction. These results highlight the effectiveness of the flexible VRF design in improving VRF efficiency.

\subsubsection{Overall PPA Comparison of FlexVector vs. Baselines}
Overall, the ablation studies evaluate different incremental optimizations, where the benefit of each depends on parameters such as tile size and buffer/VRF configurations. As a summary, multiple-buffering achieves up to 3.34$\times$ speedup, double-VRF adds up to 15\% additional speedup, vertex-cut provides up to 5.7\% additional total area reduction, and a flexible VRF region adds up to 7.4\% additional speedup. Collectively, these incremental optimizations show that FlexVector (i.e., FlexVector~(+Flexible $k$)) outperforms the GROW-like baseline in PPA under the same buffer capacity (2\,KB Dense Buffer, $m=6$), achieving $3.78{\times}$ speedup, 40.5\% lower energy, and a comparable area cost (4.7\% higher due to slightly more complex control). GROW-like$^\dagger$, with a large buffer configuration (512KB Dense Buffer and 12KB Sparse Buffer~\cite{grow}, $m=2273$), achieves a higher $1.54{\times}$ speedup than FlexVector~(+Flexible $k$), but its area increases by over $50{\times}$ and its energy is 7.2$\times$ higher, dominated by the large SRAM buffers. These trends align with the case-study results shown in Fig.~\ref{fig:ablation}(c–d), highlighting the efficiency advantage of FlexVector under practical buffer constraints.

\begin{figure}[t]
    \vspace{-0.2cm}
    \centering
    \includegraphics[width=0.99\linewidth]{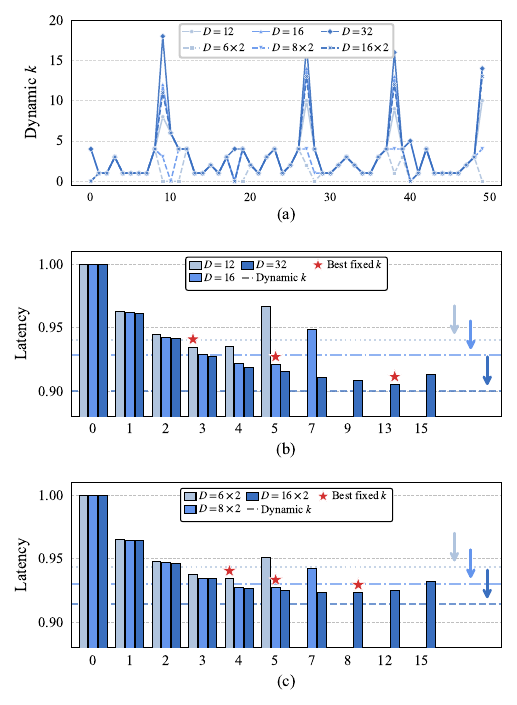}
    \caption{ Effectiveness of Algorithm~\ref{alg:topkfixed} on CiteSeer dataset under Single-VRF (VRF depth, $D \in \{12,16,32\}$) and Double-VRF ($D\in\{6{\times}2,8{\times}2,16{\times}2\}$): (a) dynamic $k$ selection across tiles; (b) latency vs.\ fixed $k$ selection (Single-VRF); (c) latency vs.\ fixed $k$ selection (Double-VRF). Dashed lines indicate the latency achieved by Algorithm~2. 
    }
    \label{fig:variesk}
    \vspace{-0.5cm}
\end{figure}

\begin{figure*}[t]
    \centering
    
    \minipage{0.49\linewidth}
        \includegraphics[width=\linewidth]{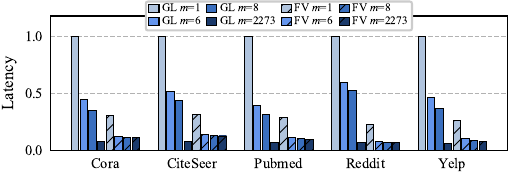}
        \vspace{-0.4cm}
        \centerline{\small (a)}
        \label{fig:latency_m}
    \endminipage
    \hfill
    \minipage{0.49\linewidth}
        \includegraphics[width=\linewidth]{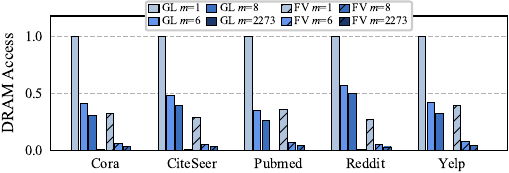}
        \vspace{-0.4cm}
        \centerline{\small (b)}
        \label{fig:dram_access}
    \endminipage

    \vspace{0.2cm}

    \minipage{0.49\linewidth}
        \includegraphics[width=\linewidth]{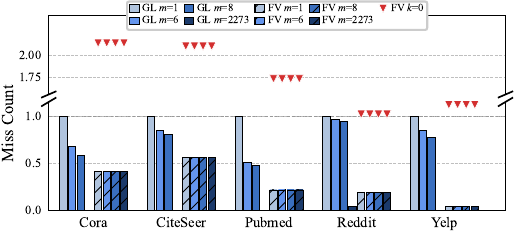}
        \vspace{-0.4cm}
        \centerline{\small (c)}
        \label{fig:miss_count}
    \endminipage
    \hfill
    \minipage{0.49\linewidth}
        \includegraphics[width=\linewidth]{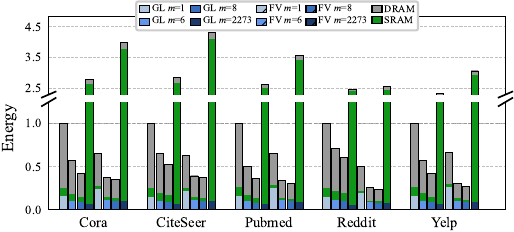}
        \vspace{-0.4cm}
        \centerline{\small (d)}
        \label{fig:isa_count}
    \endminipage
    
\caption{Comparison of GROW-like (GL) and FlexVector (FV) across varying buffer sizes $m$ on five datasets, all normalized to GL $m{=}1$: (a) latency; (b) DRAM access count; (c) miss count, with red triangles indicating FV miss count without fixed VRF region ($k{=}0$); (d) energy, where gray and green segments denote DRAM and SRAM energy.}
    \label{fig:breakdown_all}
    \vspace{-0.5cm}
\end{figure*}

\subsection{Evaluation of Flexible $k$ Selection Strategy (Algorithm~\ref{alg:topkfixed})}
\label{sec:exp-k-evalution}
Fig.~\ref{fig:variesk} evaluates Algorithm~\ref{alg:topkfixed} on the CiteSeer dataset under Single-VRF and Double-VRF configurations, where $D$ denotes the VRF depth (Single-VRF: $D \in \{12,16,32\}$; Double-VRF: $D \in \{6{\times}2,8{\times}2,16{\times}2\}$). Fig.~\ref{fig:variesk}(a) illustrates how the selected $k$, the depth of the VRF fixed region, varies across tiles during execution on FlexVector. Algorithm~\ref{alg:topkfixed} adjusts $k$ according to each tile’s sparsity, assigning larger $k$ to tiles with high dense-row reuse to maximize fixed-region utilization and smaller $k$ to tiles with dispersed accesses. Deeper VRFs consistently allow larger $k$ values.

Fig.~\ref{fig:variesk}(b–c) compares the system latency using a given $k$ (fixed across all tiles, bars) with the latency achieved by Algorithm~\ref{alg:topkfixed} (varying $k$ across tiles, dashed lines). Two key observations emerge. First, the empirically optimal $k^*$ (marked by red {$\bigstar$}) shifts across VRF depth configurations. For Single-VRF, $k^*$ grows from 3 at $D{=}12$ to 13 at $D{=}32$, as deeper VRFs can hold more fixed rows. For Double-VRF, $k^*$ increases from 4 at $D{=}6{\times}2$ to 8 at $D{=}16{\times}2$, indicating that a single fixed $k$ does not generalize across hardware settings. Second, the dashed lines in Fig.~\ref{fig:variesk} show that Algorithm~\ref{alg:topkfixed} automatically identifies a near-optimal $k$, within 2\% of the best fixed-$k$ in all cases. Overall, these results highlight that adapting the fixed–dynamic VRF partition enables efficient SpMM execution under varying sparsity and VRF depth constraints.

\subsection{Evaluation of the Impact of Various Buffer Sizes}
\label{subsec:various_buffer}

Fig.~\ref{fig:breakdown_all} compares GROW-like and FlexVector across four metrics under varying buffer sizes (e.g., Dense and Sparse Buffers), parameterized by the multi-buffer factor $m$. Larger $m$ values correspond to increased buffer capacity. For example, $m{=}6$ and $m{=}2273$ correspond to Dense Buffer sizes of 2\,KB and 512\,KB, respectively (Section~\ref{sec:Exp-baseline}).

\subsubsection{Latency}
\label{subsubsec:latency}
As shown in Fig.~\ref{fig:breakdown_all}(a), FlexVector outperforms GROW-like at $m \in \{1,6,8\}$ by up to $3.78\times$. FlexVector with $m = 1$ achieves $1.43\times$ lower latency than GROW-like with $m = 8$, indicating that FlexVector maintains higher performance even with significantly smaller buffer capacity.  However, FlexVector ($m{=}2273$) is 24.2\% slower than GROW-like ($m{=}2273$), since GROW-like's buffer-centric (e.g., cache-centric~\cite{grow}) design achieves its performance potential by progressively reducing DRAM-buffer misses as the buffer size grows, eventually reaching near-zero miss rates at $m{=}2273$. In contrast, FlexVector’s latency is primarily bottlenecked by the buffer–VRF interface (e.g., VRF data access and lane computation), and it scales more significantly with vector width than with buffer size, as discussed in Section~\ref{sec:sensitivity}.

\subsubsection{DRAM Access}
\label{subsubsec:dram_access}
Fig.~\ref{fig:breakdown_all}(b) shows that FlexVector reduces DRAM
access counts by $3.0{\times}$--$8.6{\times}$ compared to GROW-like
at the same $m$. This improvement is achieved by shifting irregular, repeated dense-row accesses from the DRAM–buffer level in GROW-like to the buffer–VRF interface, thereby avoiding DRAM accesses caused by buffer (i.e., software cache) misses. As $m$ increases, both designs reduce DRAM accesses further.
GROW-like benefits from a lower dense-row miss rate due to its larger buffer capacity, whereas FlexVector reduces per-tile DRAM traffic by amortizing burst
transfers across more tiles. At $m = 2273$, the DRAM access counts of both GROW-like and FlexVector become negligible.

\subsubsection{Dense Row Miss Count}
\label{subsubsec:cache/vrf miss}
Fig.~\ref{fig:breakdown_all}(c) compares dense-row miss counts between GROW-like (buffer/cache misses) and FlexVector (VRF misses). The miss count in GROW-like decreases monotonically with $m$, as larger buffers can capture more frequently accessed dense rows. At $m = 2273$, the expanded buffer effectively accommodates most hot dense rows, resulting in negligible misses. In contrast, FlexVector’s VRF miss count is largely independent of the buffer size ($m$), as it is governed by the VRF fixed-region size (parameterized by $k$, which is dynamically determined by Algorithm~\ref{alg:topkfixed} for each tile) rather than the buffer capacity. We further highlight the case without a VRF fixed-region ($k=0$) using red triangles. Compared to the $k>0$ case, $k=0$ leads to 3.79$\times$-27.53$\times$ higher dense-row misses, demonstrating the benefit of the fixed-region in FlexVector.

\subsubsection{Energy}
\label{subsubsec:energy}
Fig.~\ref{fig:breakdown_all}(d) shows that FlexVector achieves 13.9-63.8\% lower energy than GROW-like at the same buffer size across all datasets, for $m\in\{1,6,8\}$. At these buffer configurations, energy is dominated by DRAM accesses (gray bars), and this observation aligns with Fig.~\ref{fig:breakdown_all}(b), where FlexVector issues fewer DRAM accesses than GROW-like. At $m=2273$ with a large buffer configuration (Dense Buffer = 512KB, as in~\cite{grow}), both designs are dominated by SRAM energy (green bars), causing total energy to rise sharply. In this case, GROW-like benefits from its buffer-centric (e.g.,cache-centric~\cite{grow}) design, achieving near-zero dense-row misses, which shortens execution latency (aligned with Fig.~\ref{fig:breakdown_all}(a)) and results in 3.7-33.7\% lower overall energy than FlexVector.

\begin{figure*}[t]
    \centering
    \includegraphics[width=\textwidth]{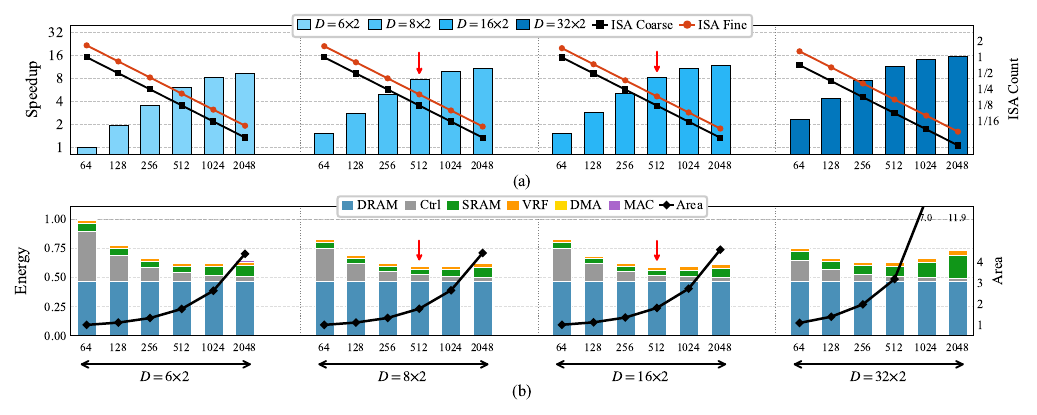}
    \caption{Impact of VRF length (VLEN = 64–2048bit) and depth ($D \in {6{\times}2,8{\times}2,16{\times}2},32{\times}2$) on FlexVector PPA: (a) speedup and instruction counts, (b) energy breakdown and area, normalized to VLEN = 64bit, $D = 6 \times 2$, with results reported as geometric means across all five datasets. $*$ Tile sizes are set to $32 \times 32$ for the first three VRF depths ($D \le 16\times2$) and $64\times64$ for the last.}
    \label{fig:vlen}
    \vspace{-0.5cm}
\end{figure*}

Overall, across different buffer configurations, GROW-like performance is highly sensitive to buffer size, with large buffers (e.g., 512KB Dense Buffer) achieving low execution latency but incurring much higher energy. In contrast, FlexVector remains largely stable with small buffers (e.g., 2KB Dense Buffer), showing no negative impact on key metrics (e.g., DRAM accesses, VRF misses). This confirms that FlexVector can deliver substantial latency and energy efficiency benefits over GROW-like under the same buffer size configurations.
 
\subsection{Evaluation of the Impact of VRF Length and Depth}
\label{sec:sensitivity}

Fig.~\ref{fig:vlen} presents the PPA evolution of FlexVector across different VRF lengths (VLEN) and depths. We first present a within-group comparison, where VLEN is varied at a fixed depth, followed by a cross-group comparison, where VRF depth is varied at a given VLEN. All results are normalized to a baseline configuration (VLEN = 64 bits, VRF depth = $6\times2$).

\subsubsection{VRF Length (Within-Group Comparison)}
\label{subsubsec:vrf_depth}
With a fixed VRF depth (e.g., $D = 6 \times 2$), increasing VLEN improves data access and computation parallelism across lanes, yielding up to a $9.44\times$ speedup and 97\% reduction in coarse-grained instructions at VLEN = 2048 bits (Fig.~\ref{fig:vlen}(a)). However, performance gains diminish beyond VLEN = 512 bits, indicating a transition from compute-bound to memory-bound execution, with DRAM bandwidth limiting throughput despite additional MAC lanes. Fig.~\ref{fig:vlen}(a) also compares FlexVector’s coarse- (black line) and fine-grained instruction counts (red line). The coarse-grained ISA executes VRF-lane data access (\texttt{MV\_Dyn}) and lane computation (\texttt{CMP}) for computing a sparse row with a dense matrix, whereas the fine-grained ISA expands these into per-nonzero and per-dense-row operations. Using coarse-grained instructions reduces total instruction count by 3–20\%, demonstrating the efficiency of our coarse-grained ISA (Section~\ref{subsec:coarseISA}). From Fig.~\ref{fig:vlen}(b), increasing VLEN increases area (by 1.76$\times$ and 4.4$\times$ at VLEN = 512 bits and 2048 bits, respectively) due to proportional scaling of MAC lanes and Dense Buffer width. System energy initially decreases to 56.2\% at VLEN = 512 bits, as higher lane parallelism reduces control overhead, but rises slightly at larger VLENs due to higher SRAM cost in the wider Dense Buffer. Similar trends across VRF depths ($D \in {6{\times}2, 8{\times}2, 16{\times}2, 32{\times}2}$) indicate that VLEN = 512 bits provides a balanced PPA tradeoff.

\subsubsection{VRF Depth (Cross-Group Comparison)}
\label{subsubsec:vrf_depth}
The cross-group comparison in Fig.~\ref{fig:vlen} shows that increasing the VRF depth to $D=32\times2$ improves system speedup to 15–135\% over $D=6\times2$ at the same VLEN, as the deeper VRF reduces misses by accommodating bigger fixed regions. Instruction count is largely independent of VRF depth ($<$1\% change from $D=6\times2$ to $D=16\times 2$) but decreases with larger tile sizes (due to coarser-grained ISA), showing a 28\% reduction when increasing the tile configuration from 32$\times$32 to 64$\times$64 at $D =32\times 2$.
Regarding area, for moderate VRF depths ($D\le16\times 2$) and VLEN ($\le 512$ bit), it remains below $2{\times}$. However, at larger depths (e.g., $D=32 \times2$) and VLEN ($\ge 1024$ bit), area grows sharply, reaching up to approximately 12$\times$, indicating that excessively wide or deep VRF configurations can incur prohibitively high area costs. Energy differences across VRF depths remain small with a given VLEN. Overall, considering both VRF length and depth, configurations with VLEN = 512 bit, specifically $D=8\times 2$ and $D=16\times2$ (highlighted by red anchors), achieve comparable PPA trade-offs, offering 6.77–7.26$\times$ speedup, 1.77–1.81$\times$ area, and 49.7–50.0\% energy reduction relative to the normalized baseline (VLEN = 64 bit, $D= 6\times2$).

\section{Conclusion}
\label{sec:conclusion}

In this paper, we present FlexVector, a hardware–software codesign for a vector-processor-based unified engine for accelerating SpMM in GCN inference. At the hardware level, FlexVector introduces software-managed flexible VRFs with fixed and dynamic regions to handle irregular data accesses at the register level, eliminating the need for large on-chip buffers or caches for repeated DRAM accesses. It also employs a coarse-grained ISA operating on sparse rows and dense submatrices to simplify SpMM control and reduce instruction count. At the software level, we propose a hybrid graph preprocessing strategy that combines inter-tile edge-cut and intra-tile vertex-cut to reshape power-law workloads under VRF capacity constraints. We further design a hierarchical dataflow that coordinates row-wise execution at the buffer–VRF level and inner-product execution at the DRAM–buffer level.

Evaluation on five real-world GCN datasets shows that FlexVector outperforms a state-of-the-art cache-centric baseline, delivering $3.78\times$ higher performance and 40.5\% lower energy at comparable area, while the sparsity-aware algorithm adapting the fixed–dynamic VRF partition achieves near-optimal performance within 2\% of the best fixed boundary. As future work, we plan to scale FlexVector by integrating multiple homogeneous vector engines to enable scalable and efficient GCN acceleration.

\section*{Acknowledgments}
This work is funded by XXX.
The authors used GPT-4o for language editing and proofreading parts of the manuscript.

\bibliographystyle{IEEEtran}  
\bibliography{references}  
\end{document}